\documentclass[letterpaper, 10 pt, conference]{ieeeconf} 

\IEEEoverridecommandlockouts                             
\usepackage{amsmath} 
\usepackage{amssymb}  

\newtheorem{mydef}{Definition}
\newtheorem{myrem}{Remark}

\usepackage{graphicx}
\usepackage{textcomp}
\usepackage{float}

\usepackage[linesnumbered,boxed,ruled,commentsnumbered]{algorithm2e}
\SetKwRepeat{Do}{do}{while}%
\usepackage{color}

\title{\LARGE \bf
Approximate Dynamic Programming for Platoon Coordination\\ 
\vspace{1.5pt}

under Hours-of-Service Regulations}

\author{Ting Bai,~Alexander Johansson,~Karl Henrik Johansson and Jonas Mårtensson
\thanks{This work is supported by the Horizon 2020 through the project ENSEMBLE, the Knut and Alice Wallenberg Foundation, the Swedish Foundation for Strategic Research and the Swedish Research Council. The work of T. Bai is also supported by the Outstanding Ph.D. Graduate Development Scholarship from Shanghai Jiao Tong University.}\vspace{1.5pt}

\thanks{T. Bai et al. are with the Integrated Transport Research Lab and Division of Decision and Control Systems, School of Electrical Engineering and Computer Science, KTH Royal Institute of Technology, Stockholm, Sweden, SE-100 44 Stockholm, Sweden. E-mails: \{{\tt\small tingbai, alexjoha, kallej, jonas1\}@kth.se}}}

\begin{document}

\maketitle
\thispagestyle{empty}
\pagestyle{empty}

\begin{abstract}
Truck drivers are required to stop and rest with a certain regularity according to the driving and rest time regulations, also called Hours-of-Service (HoS) regulations. This paper studies the problem of optimally forming platoons when considering realistic HoS regulations. In our problem, trucks have fixed routes in a transportation network and can wait at hubs along their routes to form platoons with others while fulfilling the driving and rest time constraints. We propose a distributed decision-making scheme where each truck controls its waiting times at hubs based on the predicted schedules of others. The decoupling of trucks' decision-makings contributes to an approximate dynamic programming approach for platoon coordination under HoS regulations. Finally, we perform a simulation over the Swedish road network with one thousand trucks to evaluate the achieved platooning benefits under the HoS regulations in the European Union (EU). The simulation results show that, on average, trucks drive in platoons for $37~\%$ of their routes if each truck is allowed to be delayed for $5~\%$ of its total travel time. If trucks are not allowed to be delayed, they drive in platoons for $12~\%$ of their routes. 
\end{abstract}

\section{Introduction}
Truck platooning is a technology within the field of intelligent transportation systems, allowing trucks to be linked in a line with small inter-vehicle distances and drive together. Due to the reduction of the air resistance endured by follower trucks in a platoon, truck platooning helps reduce the fuel consumption and CO$_2$ emissions. Platoon experiments have shown energy savings of $13~\%$ with $10$~m inter-vehicle gaps and energy savings of $18~\%$ with $4.7$ m gaps \cite{tsugawa2016review}. Not only is platooning technique economically and environmentally friendly but it is also beneficial to increasing the road capacity, alleviating traffic congestion and improving mobility efficiency, see, e.g., \cite{shladover2015cooperative,lioris2017platoons}.

In order for trucks with various itineraries and time schedules to fully reap the platooning benefits, efficient platoon coordination schemes are required, including, for example, path planning \cite{abdolmaleki2021itinerary}, speed regulation \cite{van2017efficient}, waiting and departure times scheduling \cite{bai2022pricing,zhang2017freight}. In this paper, we focus on the waiting time scheduling problem of truck platooning, where trucks have fixed routes in a transportation network and schedule their waiting times at hubs along their routes to facilitate the formation of platoons.   

\subsection{Motivation}
Driver fatigue has been recognized globally as one of the primary causes for traffic accidents, resulting in up to $20~\%$ of fatal road accidents \cite{bergasa2006real}, and has become a main cause for heavy truck crashes \cite{awake2011system}. To prevent driver fatigue in transport operations and increase road safety, governments of various countries impose strict regulations on trucks' driving duration and rest times, which is known as Hours-of-Service (HoS) regulations. In Table~\ref{Table1}, we give the HoS regulations of USA, EU and China, and the data is collected from \cite{jensen2009truck, poliak2018social,ChineseHosRegulations}. The maximal continuous driving time refers to the longest driving period before taking a rest, and the minimal mandatory rest time specifies the shortest time required to rest for starting a new continuous driving period according to the HoS regulations.    
\vspace{-7pt}
\begin{table}[h]
\caption{HoS regulations} 
\vspace{-8pt}
\centering 
\begin{tabular}{lccccc} 
\hline\hline 
& & & &\\[-1.5ex]
  & ~\raisebox{1.5ex}{~\textbf{USA}}~~ & \raisebox{1.5ex}{~~\textbf{EU}}~~ & ~\raisebox{1.5ex}{\textbf{China}~} 
\\ [-1.1ex]
\hline 
& & & &\\[-2ex]
Continuous driving time & & &  \\[-1.2ex]
(max.) & \raisebox{1.5ex}{$8$ h}& \raisebox{1.5ex}{$4.5$ h} & \raisebox{1.5ex}{$4$ h}\\[0.3ex]
\hline
& & & &\\[-2ex]
Mandatory rest time & & & \\[-1.5ex]
 (min.) & \raisebox{1.5ex}{$30$ min}& \raisebox{1.5ex}{$45$ min}
&\raisebox{1.5ex}{$20$ min} & \\[0.3ex]
\hline
& & & &\\[-2ex]
Daily driving time & & &  \\[-1.2ex]
(max.) & \raisebox{1.5ex}{$11$ h}& \raisebox{1.5ex}{$9$ h}
&\raisebox{1.5ex}{$10$ h} \\[0.3ex]
\hline 
\end{tabular}
\label{Table1}
\end{table}
\vspace{-3pt}

Considering the realistic HoS regulations that all trucks in the transportation system are required to follow, the scheduling of trucks' waiting times at hubs to optimally form platoons has become a new challenging problem.    

\subsection{Related Work}
Coordination of truck platoons has received considerable attention and research efforts in recent years. Existing literature has developed coordination strategies to form platoons on roads \cite{van2017fuel,liang2015heavy,larsson2015vehicle}, by means of planning routes and adjusting trucks' speeds en-route. 

An alternative to form platoons on roads is to form platoons at hubs by scheduling trucks' waiting and departure times. Most of the previous works considering hub-based platoon formation focus on coordination at a single hub \cite{boysen2018identical,larsen2019hub}, which cannot handle cases where trucks have multiple hubs along their routes. Platoon coordination at multiple hubs in a road network was considered in \cite{johansson2021real,johansson2021strategic,bai2021event}. The authors in \cite{johansson2021real} and \cite{johansson2021strategic} studied Parteo optimal and game theoretic platoon coordination solutions, respectively. In difference from our work, they do not capture realistic HoS regulations. Additionally, the decision-making of trucks in their works is coupled and the waiting times of many trucks are coordinated simultaneously, which may result in a high computation load when handling large-scale systems. In the work \cite{bai2021event}, a distributed platoon coordination approach was developed for addressing large-scale systems. This paper extends our previous work in \cite{bai2021event} by proposing an approximate dynamic programming (DP) method for platoon coordination considering realistic HoS regulations. 
\subsection{Contributions}
To the best of our knowledge, previous works on platoon coordination have not considered today's HoS regulations. The authors in \cite{xu2022truck} modeled the HoS regulations as a lower bound for the total rest time of trucks, which they allow to be distributed arbitrarily at hubs or in platoons over the whole journey. However, this is illegal according to today's HoS regulations because the rest time at a hub must exceed a certain threshold to be accounted as a mandatory rest, as shown in Table~\ref{Table1}. Motivated by the aforementioned reasons, we aim to develop a platoon coordination approach for large-scale systems that takes into account realistic HoS regulations. In our problem, each truck has a fixed route in a road network and can wait and rest at hubs along its route in order to form platoons with other trucks while respecting the HoS regulations. The contributions of this paper are summarized as follows.
\vspace{1pt}

\begin{itemize}
    \item We develop a model of the hub-based truck platoon coordination system, where the HoS regulations are included as constraints of individual trucks.
    \vspace{0.5pt}
    
    \item We propose an approximate DP approach for platoon coordination, where the decision-making of trucks is based on the predicted decisions of other trucks and decoupled. By this approach, each truck can compute its waiting times at hubs optimally while handling the HoS regulations.
    \vspace{0.5pt}
    
    \item We perform a simulation over the Swedish road network with one thousand trucks, considering the EU's HoS regulations. The simulation results show that trucks drive in platoons for $37~\%$ of their routes on average if each truck is allowed to be delayed for $5~\%$ of its total travel time. This value will be $12~\%$ if trucks are not allowed to be delayed. 
\end{itemize}
\vspace{2pt}

The rest of the paper is arranged as follows. Section~II formally presents the system model, including the road network, truck dynamics and the HoS regulations. In Section~III, the procedure employed to compute the feasible rest hubs of each truck is proposed. Section IV provides the approximate DP method for platoon coordination. In Section~V, we present the simulation conducted over the Swedish road network, followed by conclusions and future directions in Section~VI. 

\begin{figure}[t]
     \centering
     \includegraphics[width=0.96\linewidth]{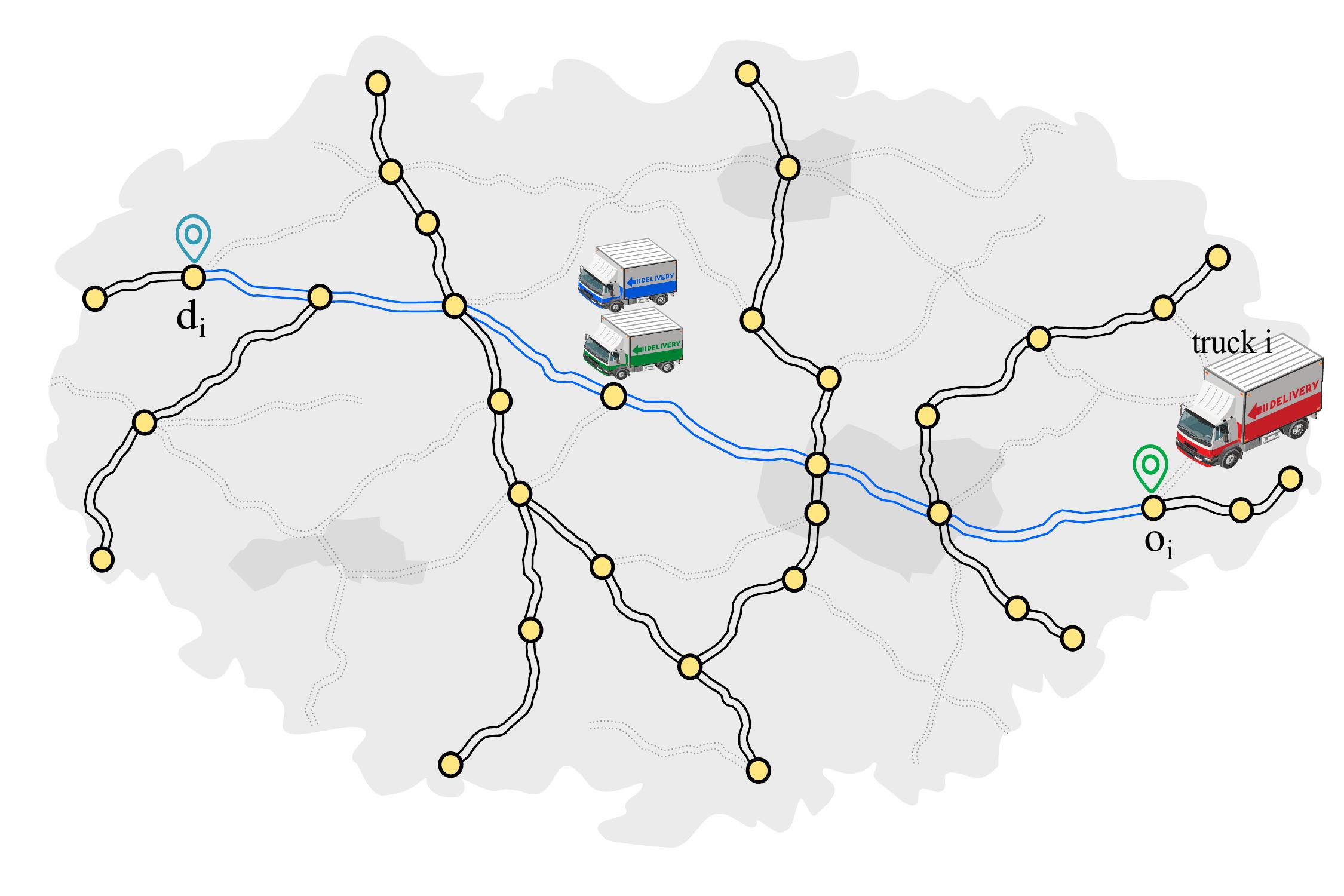}
     \vspace{-15pt}
      \caption{A transportation system with hubs (marked by yellow nodes), road segments, and trucks. The route of truck $i$ between its origin $o_i$ and destination $d_i$ is marked by the blue lines.}
      \vspace{-5pt}
      \label{Fig.1}
   \end{figure}
   
\section{System Model}
\subsection{Road Network}
We consider a large-scale transportation system illustrated in Fig.~\ref{Fig.1}, consisting of a road network with $N$ hubs and $M$ trucks from different carriers. The set of hubs and the set of trucks are denoted by $\mathcal{S}\!:=\!\{s_1,s_2,\dots,s_N\}$ and $\mathcal{M}\!:=\!\{1,2,\dots,M\}$, respectively. We assume that trucks have fixed routes in the road network, and each truck starts and ends its trip at a hub. For any truck $i\!\in\!\mathcal{M}$, the number of hubs in its route is denoted by $N_i$, and the set of its hub indices is denoted by $\mathcal{H}_i:=\{1,2,\dots,N_i\}$. The route of truck $i$ from its origin $o_i$ to its destination $d_i$ (\emph{i.e.}, between its OD-pair) is represented by the set
\begin{align}
    \mathcal{E}_i:=\{\textbf{\textit{e}}_i(1),\textbf{\textit{e}}_i(2),\dots,\textbf{\textit{e}}_i(N_i\!-\!1)\},\label{Equ.1}
\end{align}
where $\textbf{\textit{e}}_i(k)$ is the $k$-th road segment of truck $i$ connecting the $k$-th and $(k\!+\!1)$-th hub in its route. More precisely, 
\begin{align}
    \textbf{\textit{e}}_i(k):=\big(h_i(k),h_i(k\!+\!1)\big), \label{Equ.2}
\end{align}
where $h_i\!:\!\mathcal{H}_i\!\to\!{\mathcal{S}}$ denotes the map from the hub indices of truck $i$ to the set of hubs in the road network. That is, $h_i(k)$ refers to the $k$-th hub of truck $i$. Based on the above notions, the common road segment of any two trucks is defined.
\vspace{1pt}

\begin{mydef} {\label{Defition 1}}
(\textit{Common road segment}) We say that truck $i$ has its $k$-th road segment in common with truck $j$ if $\textbf{\textit{e}}_i(k)\!\in\!{\mathcal{E}_j}$, where $i,j\!\in\!\mathcal{M}$ and $i\!\neq{j}\!$.\label{Def1}
\end{mydef}

\subsection{Truck Dynamics}
With the goal of tackling the waiting time scheduling problem for truck platoon coordination under HoS regulations, the dynamics of individual trucks is presented first. The arrival time and waiting time of truck $i$ at its $k$-th hub are denoted by $a_i(k)$ and $w_i(k)$, respectively. The travel time of truck $i$ on its $k$-th road segment is denoted by $\tau_i(k)$. Then, truck $i$'s arrival time at its $(k\!+\!1)$-th hub can be computed by
\begin{align}
    a_i(k\!+\!1)=a_i(k)+w_i(k)+{\textbf{1}}_{\mathcal{H}_{i,r}}(k)t_r+\tau_i(k),\label{Equ.3}
\end{align}
where $t_r$ represents the mandatory rest time of a truck when its continuous driving time exceeds the allowed limit according to the HoS regulations. By continuous driving time, we mean the driving time without a rest. Note that, if a driver spends less time than $t_r$ at a hub, it will not be accounted as a rest and the continuous driving time will not be reset. In addition, the indicator function $\textbf{1}_{\mathcal{H}_{i,r}}$ is defined by
$\textbf{1}_{\mathcal{H}_{i,r}}:\mathcal{H}_i\!\rightarrow\!{\{0,1\}}$ where $\mathcal{H}_{i,r}\!\subset\!{\mathcal{H}_i}$
with
\begin{align}
    \textbf{1}_{\mathcal{H}_{i,r}}(k):=\begin{cases}
	1& \text{if~ $k\in{\mathcal{H}_{i,r}}$,} \\
	0& \text{if~ $k\notin{\mathcal{H}_{i,r}}$.}
	\end{cases}\label{Equ.4}
\end{align}

It is important to point out that, $\textbf{1}_{\mathcal{H}_{i,r}}(k)$ indicates whether truck $i$ takes a rest at its $k$-th hub, and both the waiting times $w_i(k)$ and the rest hubs $\mathcal{H}_{i,r}$ are our optimization variables when solving the platoon coordination problem. 

The dynamic model in Eq.~\eqref{Equ.3} applies to every truck in the transportation system, where the arrival time of truck $i$ at its origin is denoted by $t_{i,start}$. That is, $a_i(1)\!=\!t_{i,start}$. Moreover, we assume that each truck has a delivery deadline at its destination to respect while fulfilling the transport task, which requires that $a_i(N_i)\!\leq\!{t_{i,end}}$, where $t_{i,end}$ denotes truck $i$'s deadline at its destination. It is reasonable to assume that $t_{i,end}$ could cover the total travel time of truck $i$ on its routes and the rest times required at hubs. 

\subsection{Hours-of-Service Regulations}
\begin{figure}[thpb]
     \centering
     \includegraphics[width=8.3cm,height=2.9cm]{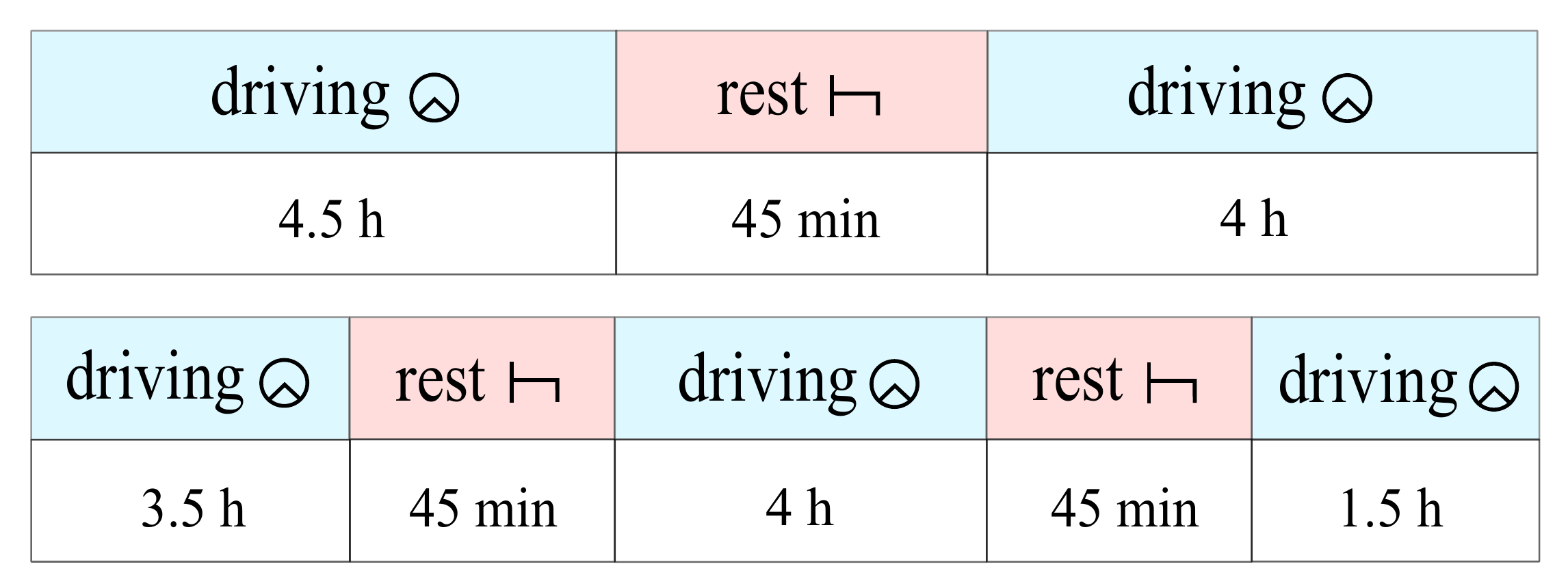}
     \vspace{-7pt}
      \caption{Two feasible driving and rest time plans according to the HoS regulations in EU, where $\bar{t}_d\!=\!4.5$~h, $t_r\!=\!45$ min and $T_d\!=\!9$ h.}
      \label{Fig.2}
   \end{figure}

We expound HoS regulations and the resulting constraints in this subsection. In line with the realistic HoS regulations in Table~\ref{Table1}, trucks are required to have a rest after a period of continuous driving. The maximum continuous driving time before taking a rest is represented by $\bar{t}_d$ and the mandatory rest time is denoted by $t_r$. Additionally, the maximum daily driving time of a truck is denoted as $T_d$. Here, we note that, in practice, the continuous driving time of a truck can be less than $\bar{t}_d$. To clarify this point, two driving and rest time plans that are feasible according to the HoS regulations in EU (see Table~\ref{Table1}) are provided in Fig.~\ref{Fig.2}.
\vspace{-9pt}
   
We assume that each truck has a route such that its travel times between hubs are less than the maximum continuous driving time, and the total travel time in its whole trip is less than the maximal daily driving time. Otherwise, trucks will not be able to complete their missions while respecting the HoS regulations. More precisely, we assume that, for any hub $k\!\in\!\{1,2,\dots,N_i\!-\!1\}$,
\begin{align}
    \tau_i(k)\leq{\bar{t}_d},~\text{and}~\sum_{k=1}^{N_i-1}\tau_i(k)\leq{T}_d.\label{Euq.5}
\end{align}

Except for the mandatory rest time at hubs, we also assume that trucks can wait extra at hubs as long as they arrive at their destinations before their deadlines. Specifically, the sum of the waiting times and rest times of any truck $i$ at all hubs in its route is constrained by
\begin{align}
    &t_{i,start}+\!\sum_{k=1}^{N_i-1}\!\big(\tau_i(k)+w_i(k)+\textbf{1}_{\mathcal{H}_{i,r}}(k)t_r\big)\leq{t_{i,end}}.\label{Equ.6}
\end{align}

The target of this paper is to optimize the waiting times $w_i(k)$ of truck $i$ and the rest hubs $\mathcal{H}_{i,r}$ at which truck $i$ should take a rest so that its own platooning benefit can be maximized while following the HoS regulations.      

\section{Feasible rest hubs}
First of all, we introduce how to determine the feasible rest hubs for each truck at which it can take a rest while meeting the HoS regulations.    

Depending on the route of a truck, it is required to rest at a number of hubs to respect the HoS regulations. To avoid serious delays at destinations, we assume that trucks will not use more rest hubs than required. In this regard, the set of feasible rest hubs $\mathcal{H}_{i,r}$ that meet the HoS regulations and minimize the number of necessary rest times is referred to as the feasible set of rest hubs of truck $i$, denoted by $\mathcal{H}_{i,r}^f$.

Given the fixed route $\mathcal{E}_i$ of any truck $i$, the travel times $\tau_i(k)$, $k\!\in\!\{1,2,\dots,N_i\!-\!1\}$ on each of its road segments, the maximum continuous driving time $\bar{t}_d$ and the mandatory rest time $t_r$, the set of feasible rest hubs of truck $i$ can be obtained by the following offline procedure.
\vspace{2pt}

\begin{itemize}
    \item [(i)] \textbf{Zero rest time}: Truck $i$ does not need to rest in its whole trip if $\sum_{k=1}^{N_i-1}\tau_i(k)\!\leq\!{\bar{t}_d}$, \emph{i.e.}, $\mathcal{H}_{i,r}^f\!=\!\emptyset$.
    \vspace{2pt}
    
    \item [(ii)] \textbf{One rest time}: If (i) is not satisfied, and there exists at least one set of a hub $\{\hat{k}\}\!\in\!\big\{\{\hat{k}\}:\hat{k}\!\in\!\{2,3,\dots,N_i\!-\!1\}\big\}$ such that 
    \begin{align}
        \sum_{k=1}^{\hat{k}-1}\tau_i(k)\leq{\bar{t}_d}~~\text{and}~\sum_{k=\hat{k}}^{N_i-1}\!\tau_i(k)\leq{\bar{t}_d}, \label{Equ.7}
    \end{align}
    then truck $i$ needs at least one rest. The set of feasible rest hubs $\mathcal{H}_{i,r}^f$ is a collection of all the sets of hubs $\{\hat{k}\}$ meeting the conditions in \eqref{Equ.7}.
    \vspace{2pt}
  
    \item [(iii)] \textbf{Two rest times}: If (i) and (ii) are not satisfied, and there exists at least one set of hubs $\{\tilde{k},\hat{k}\}\!\in\!\big\{\{k_1,k_2\}: k_1,k_2\!\in\!{\{2,3,\dots,N_i\!-\!1\}}\big\}$ with $\tilde{k}\!<\!{\hat{k}}$ such that 
    \begin{align}
        \sum_{k=1}^{\tilde{k}-1}\tau_i(k)\leq{\bar{t}_d}, ~\sum_{k=\tilde{k}}^{\hat{k}-1}\tau_i(k)\leq{\bar{t}_d},~\text{and} \sum_{k=\hat{k}}^{N_i-1}\!\tau_i(k)\leq{\bar{t}_d},\label{Equ.8}
    \end{align}
    then truck $i$ needs at least two rests. The set of feasible rest hubs $\mathcal{H}_{i,r}^f$ is a collection of all the sets of hubs $\{\tilde{k},\hat{k}\}$ meeting the conditions in \eqref{Equ.8}.  
\end{itemize}
\vspace{1pt}

\begin{myrem}
In the above procedure for determining $\mathcal{H}_{i,r}^f$, we disregard the other cases with more than two rest times. However, the proposed method can be easily extended to handle those cases. From the perspective of practical applications, two rest times are enough to address most realistic scenarios under the HoS regulations in Table~\ref{Table1}.\label{Rem1}
\end{myrem}

\begin{myrem}
Note that $\mathcal{H}_{i,r}$ is one of the elements selected from the set of feasible rest hubs $\mathcal{H}_{i,r}^f$, and each element in $\mathcal{H}_{i,r}^f$ satisfies HoS regulations. If $\mathcal{H}_{i,r}^f\!\neq\!{\emptyset}$, then each element in $\mathcal{H}_{i,r}^f$ is a set and has the same cardinality. The cardinality of each element (\emph{i.e.}, a set) in $\mathcal{H}_{i,r}^f$ equals the least number of rest times that truck $i$ needs to meet HoS regulations. \label{Rem2}
\end{myrem}

\section{Approximate Dynamic Programming for Platoon Coordination}
This section presents an approximate dynamic programming approach for addressing the platoon coordination problem of individual trucks under HoS regulations. We start by introducing the utility of each truck. 

\subsection{Utility}
Trucks can decide their waiting times at hubs along their routes in order to form platoons with others and enjoy the platooning benefit. The time that trucks wait at a hub relies on the predicted platooning reward they may achieve by joining a platoon and the predicted loss caused by their waiting decisions. That is, the predicted platooning reward and the predicted waiting loss.       
\vspace{3pt}

\subsubsection{Predicted Platooning Reward} For any truck $i$ arriving at its $k$-th hub, it cares about the total platooning reward it predicts to obtain from its current hub $k$ to its destination $d_i$. We denote by $\textbf{\textit{w}}_i(k)$ the waiting times of truck $i$ at its $k$-th hub, including its waiting time at the current hub $k$ and its predicted waiting times at the remaining hubs $(k\!+\!h)$ with $h\!\in\!\{1,\dots,N_i\!-\!1\!-\!k\}$, which has the form of
\begin{align}
    \textit{\textbf{w}}_i(k):=[w_i(k|k),w_i(k\!+\!1|k),\dots,w_i(N_i\!-\!1|k)],\label{Equ.9}
\end{align}
where $w_i(k\!+\!h|k)$ denotes the predicted waiting time of truck $i$ at its $(k\!+\!h)$-th hub predicted at its $k$-th hub. Given the routes and predicted departure times of other trucks, truck $i$ is able to predict its platooning reward on the road segments $\{\textbf{\textit{e}}_i(k),\textbf{\textit{e}}_i(k\!+\!1),\dots,\textbf{\textit{e}}_i(N_i\!-\!1)\}$. 
\vspace{2pt}

In what follows, the predicted platooning reward on one road segment $\textbf{\textit{e}}_i(k\!+\!h)$ of truck $i$ is given. Let $\mathcal{P}_i(k\!+\!h|k)$ be the predicted platooning partners of truck $i$ at its $(k\!+\!h)$-th hub predicted at its $k$-th hub, which is a set of trucks that are predicted by truck $i$ to form a platoon with truck $i$. Mathematically, $\mathcal{P}_i(k\!+\!h|k)$ is represented as 
\begin{align}
    \mathcal{P}_i(k\!+\!h|k)=\big\{j:~&j\!\in\!\mathcal{M},~\textbf{\textit{e}}_i(k\!+\!h)\!\in\!\mathcal{E}_j,~\text{and}\nonumber\\
    &d_i(k\!+\!h|k)=d_j^{i,k}\big(h_i(k\!+\!h)\big)\big\}, \label{Equ.10}
\end{align}
where $\textbf{\textit{e}}_i(k+h)\!\in\!\mathcal{E}_j$ indicates that truck $i$ has its $(k\!+\!h)$-th road segment in common with truck $j$. Meanwhile, $d_i(k\!+\!h|k)=d_j^{i,k}\big(h_i(k\!+\!h)\big)$ requires that truck $i$ and $j$ departure from the $(k\!+\!h)$-th hub of truck $i$ at the same time, where $d_j^{i,k}\big(h_i(k\!+\!h)\big)$ is the predicted departure time of truck $j$ at the $(k\!+\!h)$-th hub of truck $i$, known by truck $i$ when it arrives at its $k$-th hub. Moreover, $d_i(k+h|k)$ denotes the predicted departure time of truck $i$ itself at its $(k\!+\!h)$-th hub, which is determined by
\begin{align}
    &d_i(k\!+\!h|k)\nonumber\\
    &=a_i(k\!+\!h|k)+w_i(k\!+\!h|k)+\textbf{1}_{\mathcal{H}_{i,r}(k)}(k\!+\!h)t_r,\label{Equ.11}
\end{align}
where $a_i(k\!+\!h|k)$ and $w_i(k\!+\!h|k)$ represent the predicted arrival and waiting times of truck $i$ at its $(k\!+\!h)$-th hub predicted at its $k$-th hub, respectively. It is worth noting that, in Eq.~\eqref{Equ.11}, $\mathcal{H}_{i,r}(k)\!\in\!{\mathcal{H}_{i,r}^f}$ denotes the set of rest hubs of truck $i$ determined at its $k$-th hub. In line with the indicator function defined in \eqref{Equ.4}, we have that 
\begin{align}
    \textbf{1}_{\mathcal{H}_{i,r}(k)}(k\!+\!h)=\begin{cases}
	1& \text{if~ $k\!+\!h\in{\mathcal{H}_{i,r}}(k)$,} \\
	0& \text{if~ $k\!+\!h\notin{\mathcal{H}_{i,r}}(k)$.}
	\end{cases}\label{Equ.12}
\end{align}
Recall that $t_r$ is the mandatory rest time by HoS regulations. Therefore, $\textbf{1}_{\mathcal{H}_{i,r}(k)}(k\!+\!h)t_r$ depicts the rest time of truck $i$ at its $(k\!+\!h)$-th hub. The conditions in Eq. \eqref{Equ.10} ensure that truck $i$ and $j$ can form a platoon on the road segment $\textbf{\textit{e}}_i(k\!+\!h)$ in the prediction of truck $i$ at its $k$-th hub.   
\vspace{1pt}

In the sequel, we use $p_i(k\!+\!h|k)\!=\!|\mathcal{P}_i(k\!+\!h|k)|$ to denote the cardinality (or the \textit{size}) of the set $\mathcal{P}_i(k\!+\!h|k)$, \emph{i.e.}, the number of trucks included in $\mathcal{P}_i(k\!+\!h|k)$, including truck $i$ itself. The predicted platooning reward of truck $i$ on its road segment $\textbf{\textit{e}}_i(k\!+\!h)$ predicted at its hub $k$ can be defined as
\begin{align}
    R_i(k\!+\!h|k):=\xi_i\tau_i(k\!+\!h)\frac{p_i(k\!+\!h|k)\!-\!1}{p_i(k\!+\!h|k)},\label{Equ.13}
\end{align}
where $\xi_i$ denotes the monetary platooning benefit from fuel savings per follower truck and per time unit, and $\tau_i(k\!+\!h)$ is the travel time of truck $i$ on its $(k\!+\!h)$-th road segment. As each follower truck in a platoon saves approximately the same fuel while the leader truck has a significantly smaller fuel saving, according to the platoon field experiments \cite{davila2013environmental,bishop2017evaluation}, we assume that the total platooning benefit is evenly shared among the $p_i(k\!+\!h|k)$ trucks in Eq.~\eqref{Equ.13}. 
\vspace{1pt}

Consequently, the platooning reward that truck $i$ predicts to achieve from its $k$-th hub to the end of its trip is 
\begin{align}
    R_i(k)=\sum_{h=0}^{N_i-1-k}R_i(k\!+\!h|k).\label{Equ.14}
\end{align}

\subsubsection{Predicted Waiting Loss}
The decision to wait at hubs could also lead to some benefit loss for truck $i$ due to a higher labor cost and risk for delay. We denote by $L_i(k\!+\!h|k)$ the predicted waiting loss of truck $i$ at its $(k\!+\!h)$-th hub predicted at its $k$-th hub, which is defined by
\begin{equation}
    L_i(k\!+\!h|k):=\epsilon_iw_i(k\!+\!h|k),\label{Equ.15}
\end{equation}
where $\epsilon_i$ denotes the monetary loss of truck $i$ per time unit for waiting. The predicted waiting loss at all hubs between the $k$-th and $(N_i\!-\!1)$-th hub of truck is then denoted as 
\begin{align}
    L_i(k)=\sum_{h=0}^{N_i-1-k}L_i(k\!+\!h|k).\label{Equ.16}
\end{align}

Given the above, the utility of each truck $i$ at its $k$-th hub is represented as 
\begin{align}
    J_i(k)=R_i(k)-L_i(k).\label{Equ.17}
\end{align}

\begin{myrem}
With the knowledge of the other trucks' routes and time schedules, truck $i$ is capable of planning its own waiting times $\textbf{\textit{w}}_i(k)$ and its rest hubs $\mathcal{H}_{i,r}(k)$ for achieving a maximized utility.\label{Rem3}
\end{myrem}

\subsection{Distributed Platoon Coordination}
Next, we will introduce how trucks decide their optimal $\textbf{\textit{w}}_i(k)$ and $\mathcal{H}_{i,r}(k)$ by an approximate DP approach. Recall that, the feasible set of rest hubs $\mathcal{H}_{i,r}^f$ computed offline provides all the feasible solutions to $\mathcal{H}_{i,r}(k)$ with each solution meeting the HoS regulations. We denote by $\mathcal{D}_{i,r}(k)$ the set of hubs at which truck $i$ has taken a rest when arriving at its $k$-th hub, which is initialized and updated by
\begin{align}
    &\mathcal{D}_{i,r}(1)=\emptyset\label{Equ.18}\\
   &\mathcal{D}_{i,r}(k\!+\!1)=\mathcal{D}_{i,r}(k)\!\cup\!\{k\},~\text{if}~k\!\in\!{\mathcal{H}_{i,r}^*}(k),\label{Equ.19}
\end{align}
where $\mathcal{H}_{i,r}^*(k)$ is the optimal $\mathcal{H}_{i,r}(k)$ that maximizes truck $i$'s utility $J_i(k)$, which can be obtained by solving the following problem \eqref{Equ.22}. Accordingly, we denote by $\tilde{\mathcal{H}}_{i,r}^f(k)$ the set of feasible rest hubs of truck $i$ at its $k$-th hub, which is dynamically updated in line with $\mathcal{D}_{i,r}(k)$, that is
\begin{align}
&\tilde{\mathcal{H}}_{i,r}^f(1)=\mathcal{H}_{i,r}^f,\label{Equ.20}\\
   & \tilde{\mathcal{H}}_{i,r}^f(k\!+\!1)\nonumber\\
   &=\big\{\mathcal{H}_{i,r}\!\in\!\tilde{\mathcal{H}}^{f}_{i,r}(k):k^{'}\!\!\in\!\mathcal{H}_{i,r}~\text{if}~k^{'}\!\!\in\!{\mathcal{D}_{i,r}(k\!+\!1)}\big\}.\label{Equ.21}
\end{align}
Notice that, $\tilde{\mathcal{H}}_{i,r}(k\!+\!1)$ is comprised of all the elements in $\tilde{\mathcal{H}}_{i,r}(k)$ that cover the selected rest hubs in $\mathcal{D}_{i,r}(k\!+\!1)$. For each truck $i$, it is necessary to update its feasible rest hubs according to \eqref{Equ.21} because not all the elements in the initial feasible set $\mathcal{H}_{i,r}^f$ are feasible anymore when some rest hubs of truck $i$ have been used.   
\vspace{1pt}

Based on the above descriptions and definitions, the optimal waiting times and rest hubs of truck $i$ at its $k$-th hub can be computed by solving the following distributed platoon coordination problem 
\begin{flalign}
    &\max_{\substack{{\textbf{\textit{w}}_i(k)}},\mathcal{H}_{i,r}(k)}~J_i(k)&\label{Equ.22}
\end{flalign}
\vspace{-10pt}
\begin{subequations}
    \begin{align}
    \!\!\!\mathrm{s.t.}~~&a_i(k|k)\!=\!t_{i,arr}(k)\tag{\ref{Equ.22}a}\\
    &a_i(k\!+\!h\!+\!1|k)\!=\!a_i(k\!+\!h|k)\!+\!w_i(k\!+\!h|k)\!+\!\tau_i(k\!+\!h)\nonumber\\
    &\quad +\!\textbf{1}_{\mathcal{H}_{i,r}(k)}(k\!+\!h)t_r,~h\!=\!0,1,\dots,N_i\!-\!1\!-\!k\tag{\ref{Equ.22}b}\\
    &\mathcal{H}_{i,r}(k)\in{\tilde{\mathcal{H}}_{i,r}^f(k)}\tag{\ref{Equ.22}c}\\
    &a_i(N_i|k)\leq{t_{i,end}},\tag{\ref{Equ.22}d}
\end{align}
\end{subequations}
where $t_{i,arr}(k)$ denotes the arrival time of truck $i$ at its $k$-th hub with $t_{i,arr}(1)\!=\!t_{i,start}$. The set of feasible rest hubs $\tilde{\mathcal{H}}_{i,r}^f(k)$ of truck $i$ at its $k$-th hub is initialized by \eqref{Equ.20} and updated via \eqref{Equ.19} and \eqref{Equ.21}. The constraint (\ref{Equ.22}d) implies that each truck respects its deadline at the destination.   
\vspace{2.5pt}

\begin{myrem}
The problem \eqref{Equ.22} is a mixed integer nonlinear programming (MINP) problem. In the proposed distributed platoon coordination scheme, trucks compute their optimal waiting and rest decisions $\textbf{\textit{w}}_i^*(k)$ and $\mathcal{H}_{i,r}^*(k)$ only once when arriving at hubs, based on the predicted schedules of other trucks. Such a decoupling of trucks' decision-makings allows us to solve the problem \eqref{Equ.22} via approximate DP \cite{bertsekas2019reinforcement}.\label{Rem4}
\end{myrem}
\vspace{2pt}

\begin{myrem}
Truck $i$ takes a rest at its hub $k$ if $k\!\in\!{\mathcal{H}_{i,r}^*(k)}$. Meanwhile, its optimal waiting time implemented at its current hub $k$ is $w_i^*(k)$ (\emph{i.e.}, the first item of $\textbf{\textit{w}}_i^*(k)$). The other rest hubs in $\mathcal{H}_{i,r}^*(k)$ and the other waiting times in $\textbf{\textit{w}}_i(k)$ will serve as the predicted schedules of truck $i$ for other trucks to compute their optimal decisions.\label{Rem5}
\end{myrem}

\section{Simulation over Swedish Road Network}
\begin{figure}[t]
     \centering
     \includegraphics[width=0.445\linewidth]{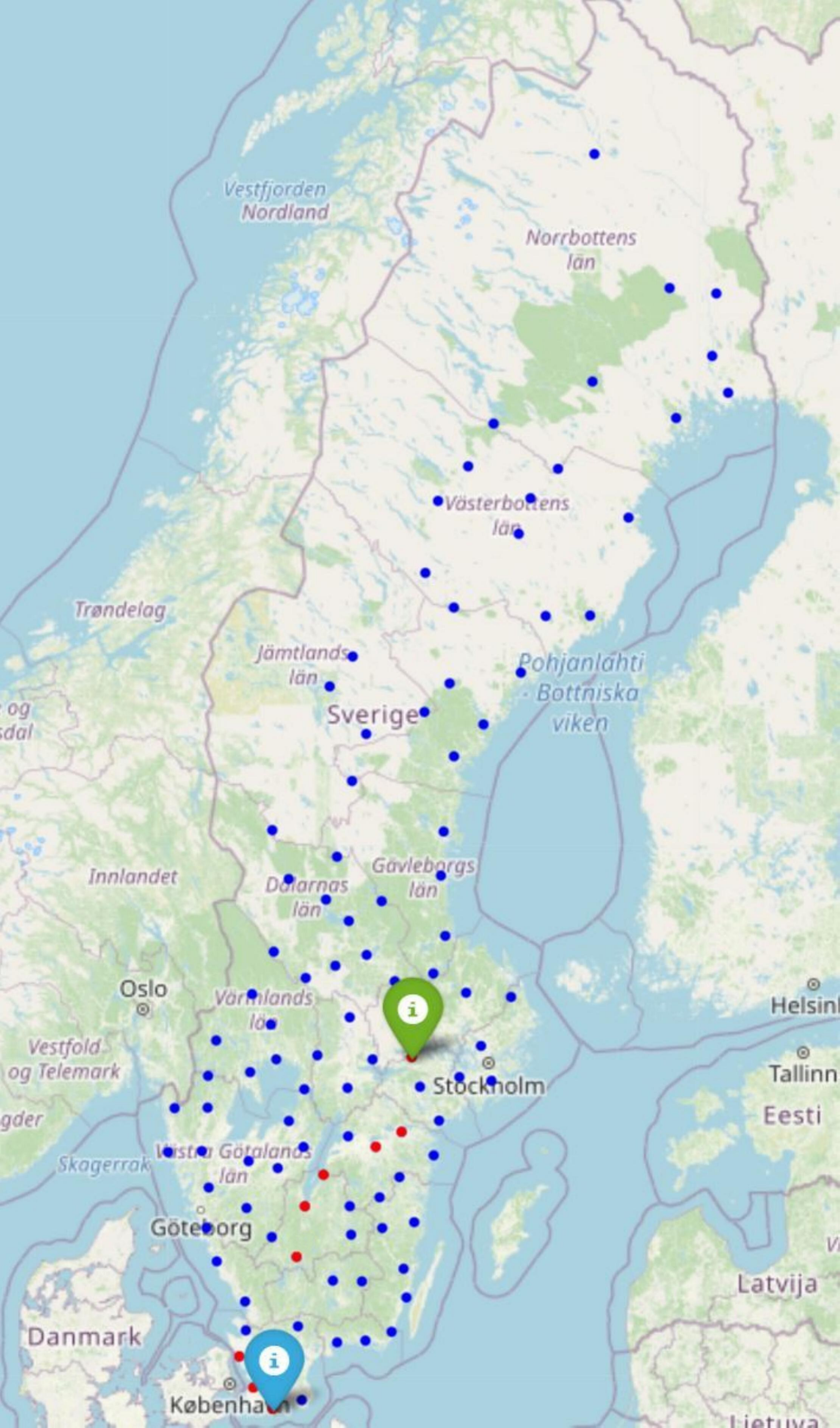}
     \vspace{-2pt}
      \caption{The Swedish road network with $105$ major hubs (marked by blue nodes), where hubs along the route of one truck are marked by red nodes.}
      \vspace{-3pt}
      \label{Fig.3}
   \end{figure}

This section performs a simulation over the Swedish road network to illustrate the effectiveness of the developed platoon coordination method. The considered transportation network is shown in Fig.~\ref{Fig.3}, which includes $105$ hubs in Sweden. The hubs are real road terminals selected from the SAMGODS model \cite{bergquist2016representation}, which is the national model for freight transportation in Sweden. The HoS regulations in EU in Table~\ref{Table1} are considered in the simulation.  

In the simulation, we consider $1000$ trucks and the OD-pair of each truck is randomly generated from the $105$ hubs such that the probability of selecting an OD-pair is proportional to its truck flow in the SAMGODS model. The routes of the trucks are obtained from \textit{OpenStreetMap}~\cite{OpenStreetMap}. 

Other simulation settings are given as follows. We assume that trucks start their trips at a random time between 08:00--10:00. The delivery deadline of each truck is set such that its total waiting time over its route (also called waiting budget) is less than $5~\%$ of its total travel time. The speed of each truck is $80$ km/h. The fuel price is around $1.73$\texteuro ~per liter and the fuel consumption of each follower truck is assumed to be reduced by $10~\%$. Moreover, the platooning benefit is $\xi_i\!=\!5.5$\texteuro ~per follower truck per hour, and the waiting loss of each truck is $\epsilon_i\!=\!25$\texteuro ~per hour. 

\begin{table}[t]
\caption{The number of rest times required for trucks} 
\vspace{-7pt}
\centering 
\begin{tabular}{l|c|cc|cc} 
\hline 
& & \multicolumn{2}{c}{}  & \multicolumn{2}{|c}{}\\[-1.5ex]
   & \raisebox{1.5ex}{\textbf{Zero rest time}} & \multicolumn{2}{c}{\raisebox{1.5ex}{\textbf{One rest time}}} & \multicolumn{2}{|c}{\raisebox{1.5ex}{\textbf{Two rest times}}} 
\\ [-1.1ex]
\hline 
 & & \multicolumn{2}{c}{} & \multicolumn{2}{|c}{}\\[-0.7ex]
\raisebox{1.5ex}{Nr. of trucks} & \raisebox{1.5ex}{$706$}& \multicolumn{2}{c}{\raisebox{1.5ex}{$250$}} & \multicolumn{2}{|c}{\raisebox{1.5ex}{$44$}}\\[-0.4ex]
\hline
& & \multicolumn{2}{c}{}  & \multicolumn{2}{|c}{} \\[-1.2ex]
\raisebox{1.5ex}{Size of $\mathcal{H}_{i,r}^f$} & \raisebox{1.5ex}{$0$} & ~\raisebox{1.5ex}{$=\!1$} & \raisebox{1.5ex}{$>\!1$}~
&~\raisebox{1.5ex}{$=\!1$} & \raisebox{1.5ex}{$>\!1$}~ \\[-0.4ex]
\hline 
& & \multicolumn{2}{c}{}  & \multicolumn{2}{|c}{} \\[-0.5ex]
 \raisebox{1.5ex}{Nr. of trucks} & \raisebox{1.5ex}{$706$} & ~\raisebox{1.5ex}{$113$} & \raisebox{1.5ex}{$137$}~
&~\raisebox{1.5ex}{$2$} & \raisebox{1.5ex}{$42$}~ \\[-0.1ex]
\hline 
\end{tabular}
\label{Table2}
\end{table}

Table~\ref{Table2} shows how many trucks that are required to take zero, one and two mandatory rests according to their routes in the simulation. It also shows how many trucks that have only one feasible option and two or more feasible options of the rest hubs that meet HoS regulations, where the size of $\mathcal{H}_{i,r}^f$ equals the number of feasible options for the rest hubs. As we can see, $706$ trucks are not required to rest since their total travel times are less than the maximal continuous driving period. It can be also seen that $250$ and $44$ trucks are required to rest once and twice, respectively. Out of the trucks that are required to rest once and twice, $137$ and $42$ trucks have more than one feasible option for the rest hubs.

\begin{figure}[t]
     \centering
     \includegraphics[width=0.97\linewidth]{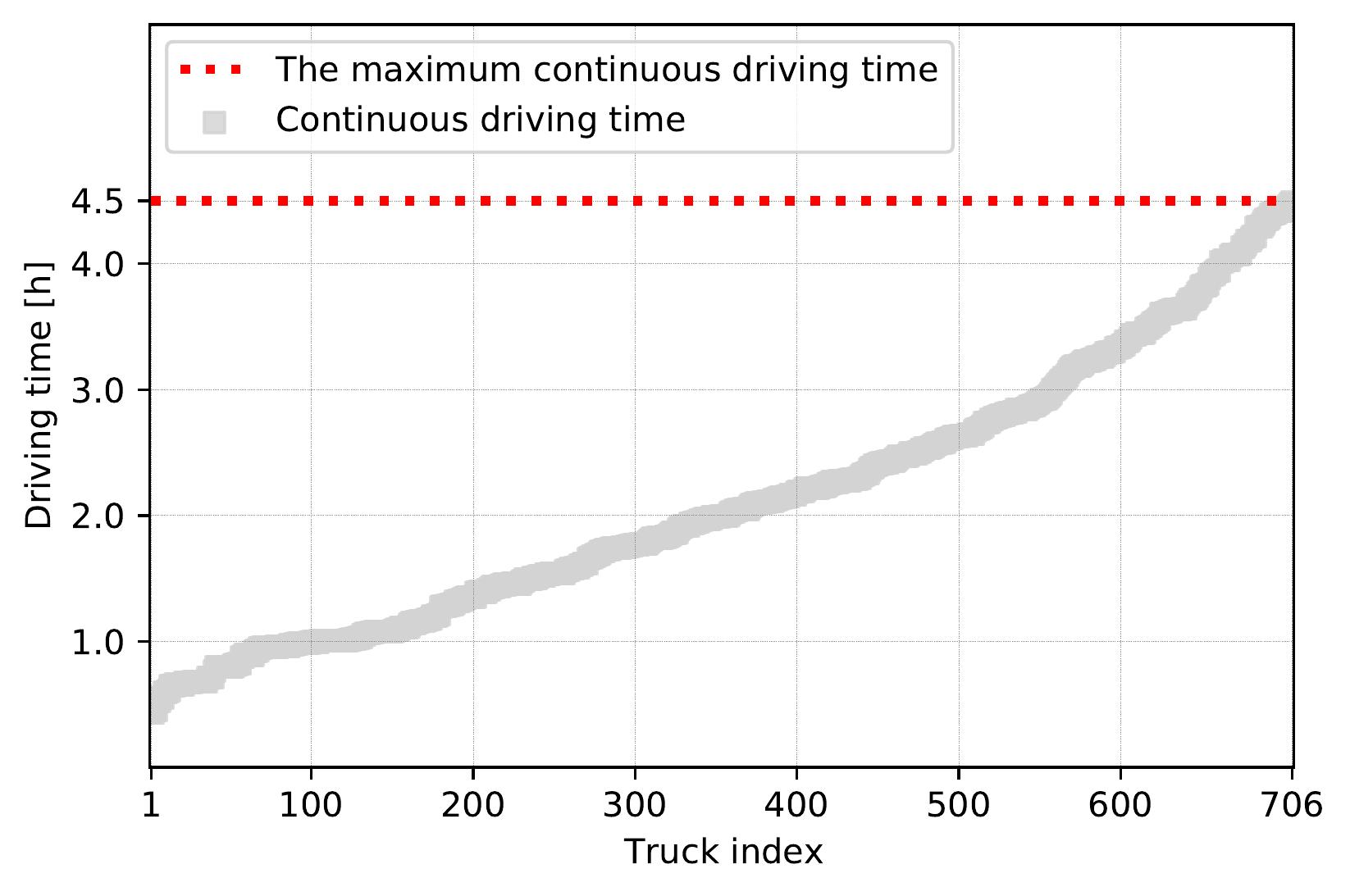}
     \vspace{-5pt}
      \caption{Continuous driving time of each truck with zero rest time.}
      \label{Fig.4}
\end{figure}

\begin{figure}[t]
     \centering
     \includegraphics[width=0.97\linewidth]{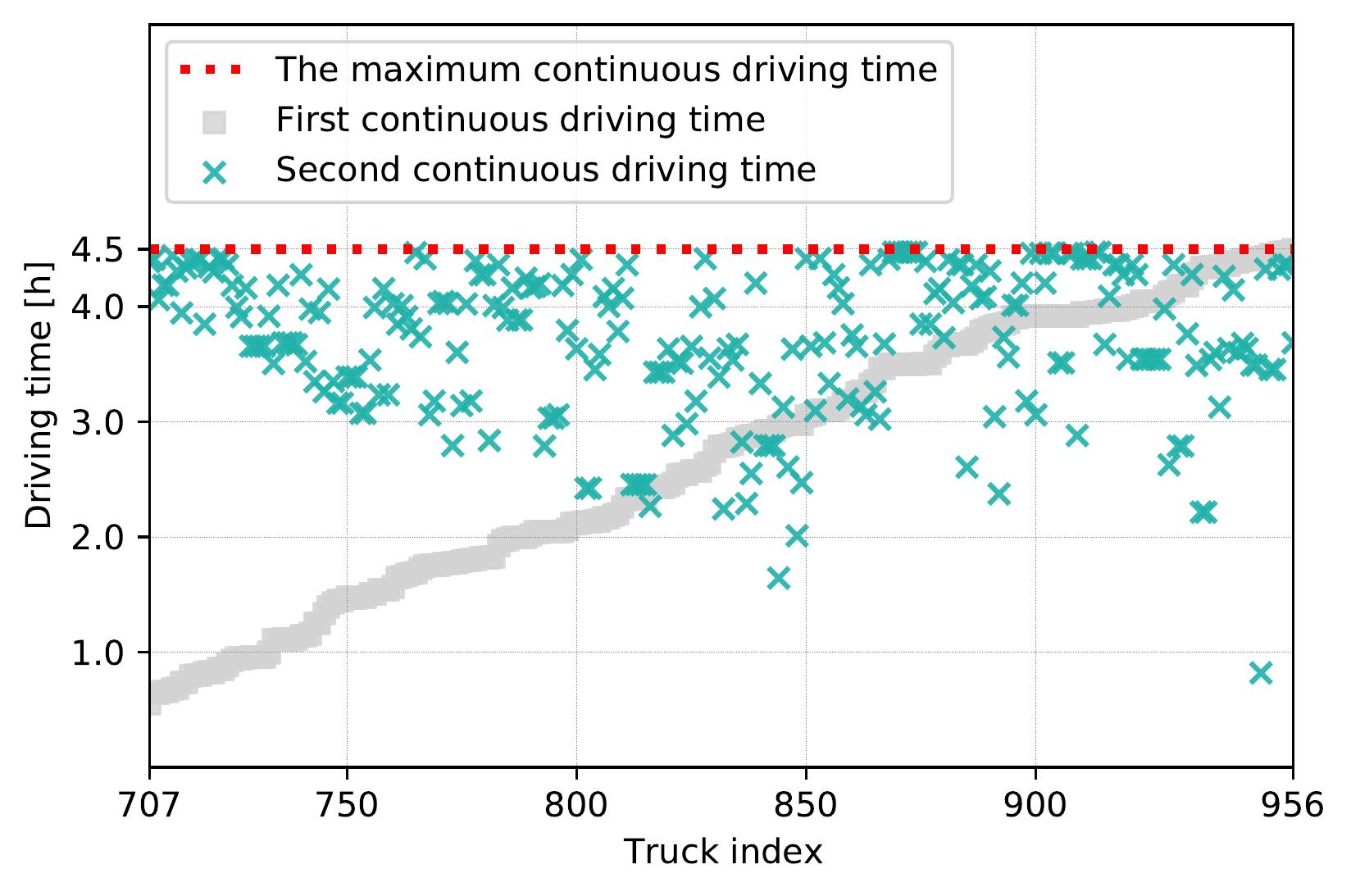}
     \vspace{-5pt}
      \caption{Continuous driving time of each truck with one rest time.}
      \label{Fig.5}
\end{figure}

\begin{figure}[t]
     \centering
     \includegraphics[width=0.975\linewidth]{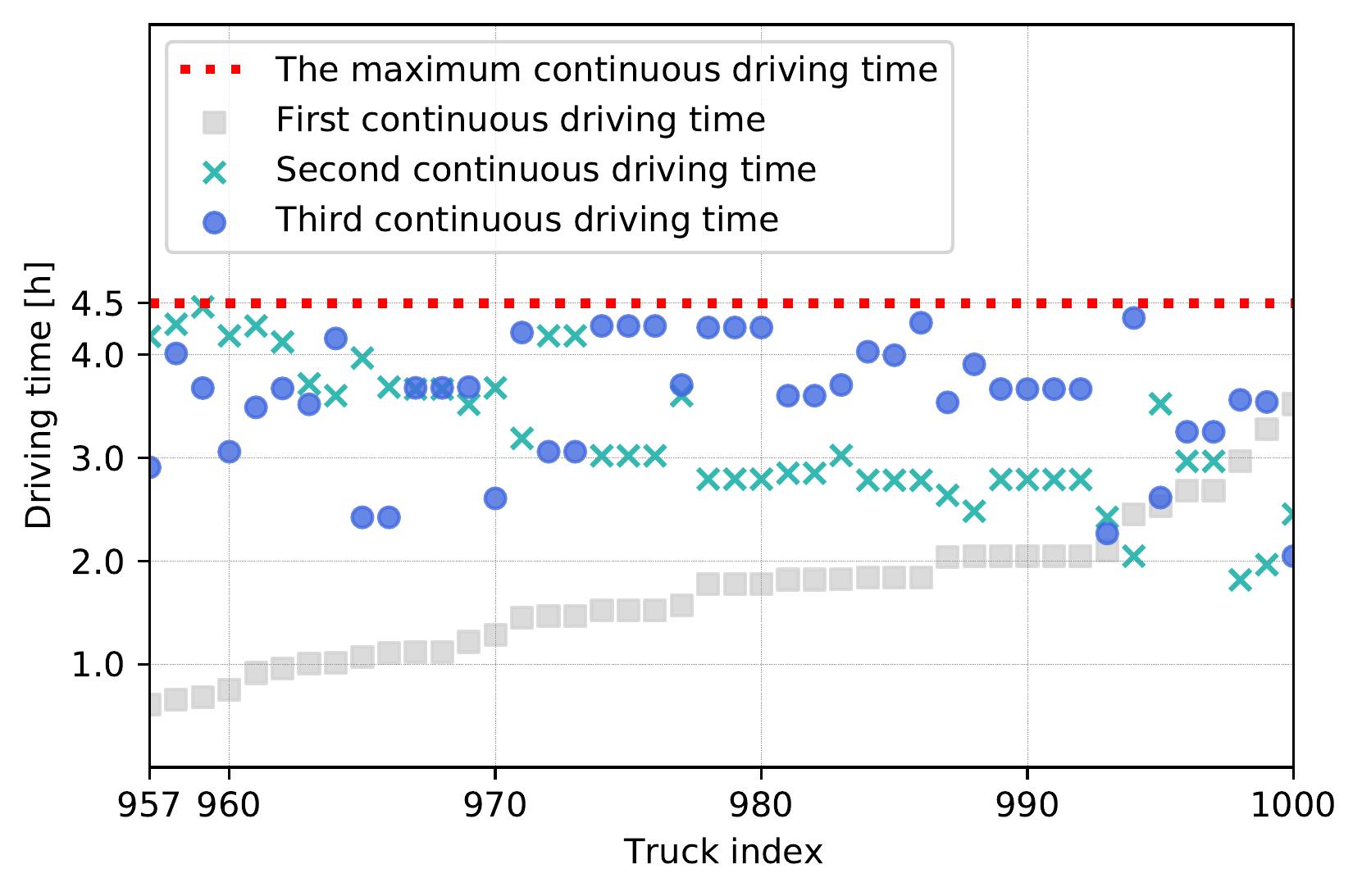}
     \vspace{-5pt}
      \caption{Continuous driving time of each truck with two rest times.}
      \label{Fig.6}
\end{figure}

Figs.~\ref{Fig.4}--\ref{Fig.6} show the continuous driving times of each truck between its optimal rest hubs, where the truck indices are sorted according to their first continuous driving time. Specifically, Fig.~\ref{Fig.4} gives the driving times of trucks that are not required to rest. Figs.~\ref{Fig.5} and \ref{Fig.6} show the continuous driving times of trucks that are required to rest once and twice, respectively. From the figures we can see, all the continuous driving times of trucks are less than $4.5$ hours, and hence, meeting the EU's HoS regulations. Furthermore, the result in Fig.~\ref{Fig.5} indicates that the shorter the first continuous driving period is, the longer is the second continuous driving period. Such a trend can also be seen in Fig.~\ref{Fig.6}.

\begin{figure}[t]
     \centering
     \includegraphics[width=0.973\linewidth]{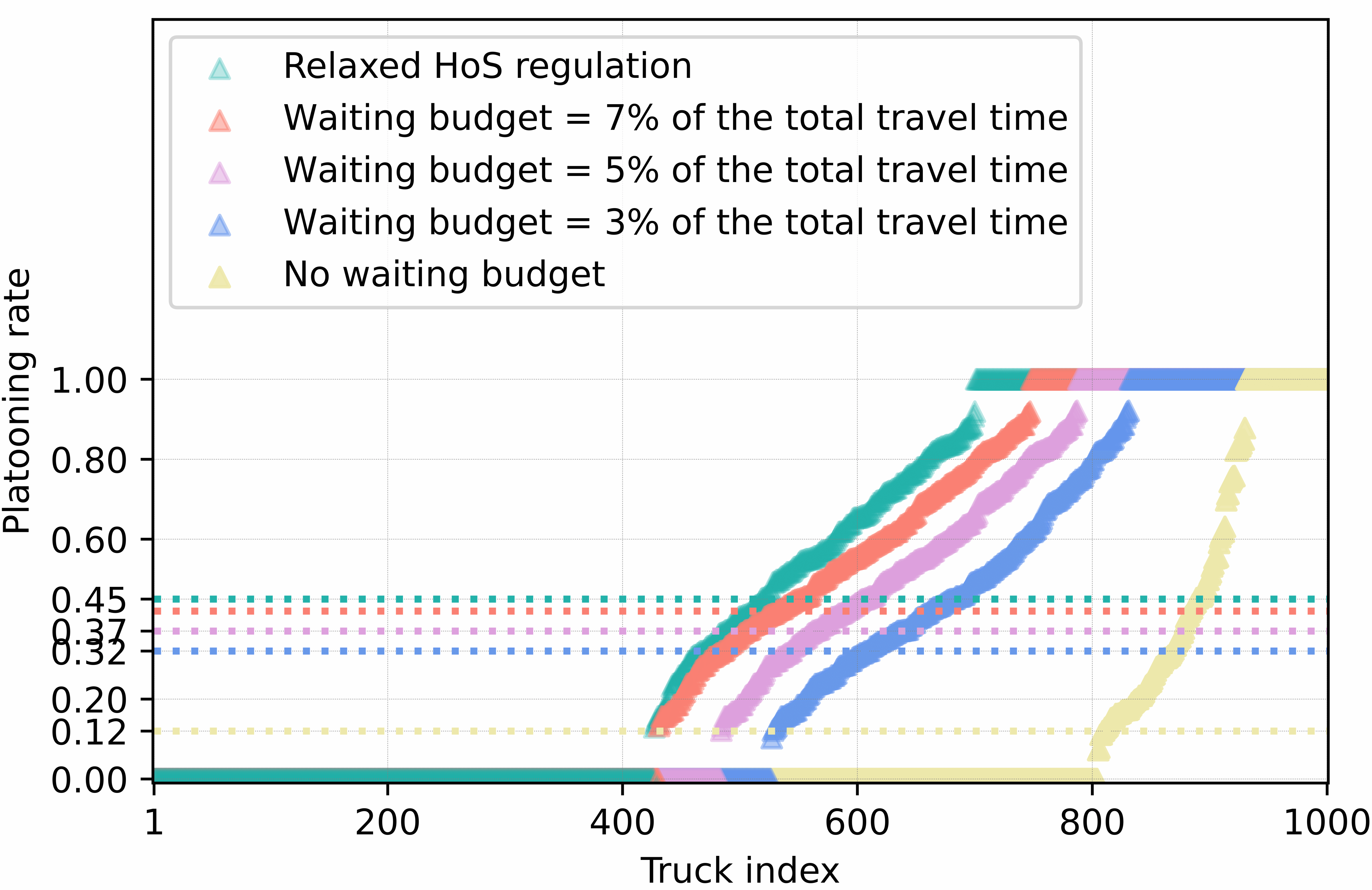}
     \vspace{-2.3pt}
      \caption{The platooning rate of each truck and the average platooning rate (dotted lines) for different waiting budgets.}
      \label{Fig.7}
\end{figure}

The coordination efficiency of the developed method is further evaluated by the platooning rate, which is defined for each truck $i$ as its total travel time in platoons over its total travel time in the network. In Fig.~\ref{Fig.7}, we provide the platooning rate of each truck under different settings of the waiting budget, where the truck indices are resorted according to their platooning rates. The average platooning rates with the waiting budgets of $0~\%$, $3~\%$, $5~\%$ and $7~\%$ of the total travel times are $0.12$, $0.32$, $0.37$, $0.42$, respectively. This result shows that the waiting budget has a large impact on the platooning rate. The higher the waiting budget is, the higher is the average platooning rate. We also compare with a relaxed HoS regulation case, where trucks have a waiting budget of $7~\%$ of their total travel times and, in addition, can spend their mandatory rest times arbitrarily along their journeys, as in \cite{xu2022truck}. Fig.~\ref{Fig.7} shows that the case with relaxed HoS yields the highest platooning rate. 
For the case with no waiting budget in Fig.~\ref{Fig.7}, only the set of rest hubs is optimized to form platoons.

\begin{figure}[t]
     \centering
     \includegraphics[width=0.975\linewidth]{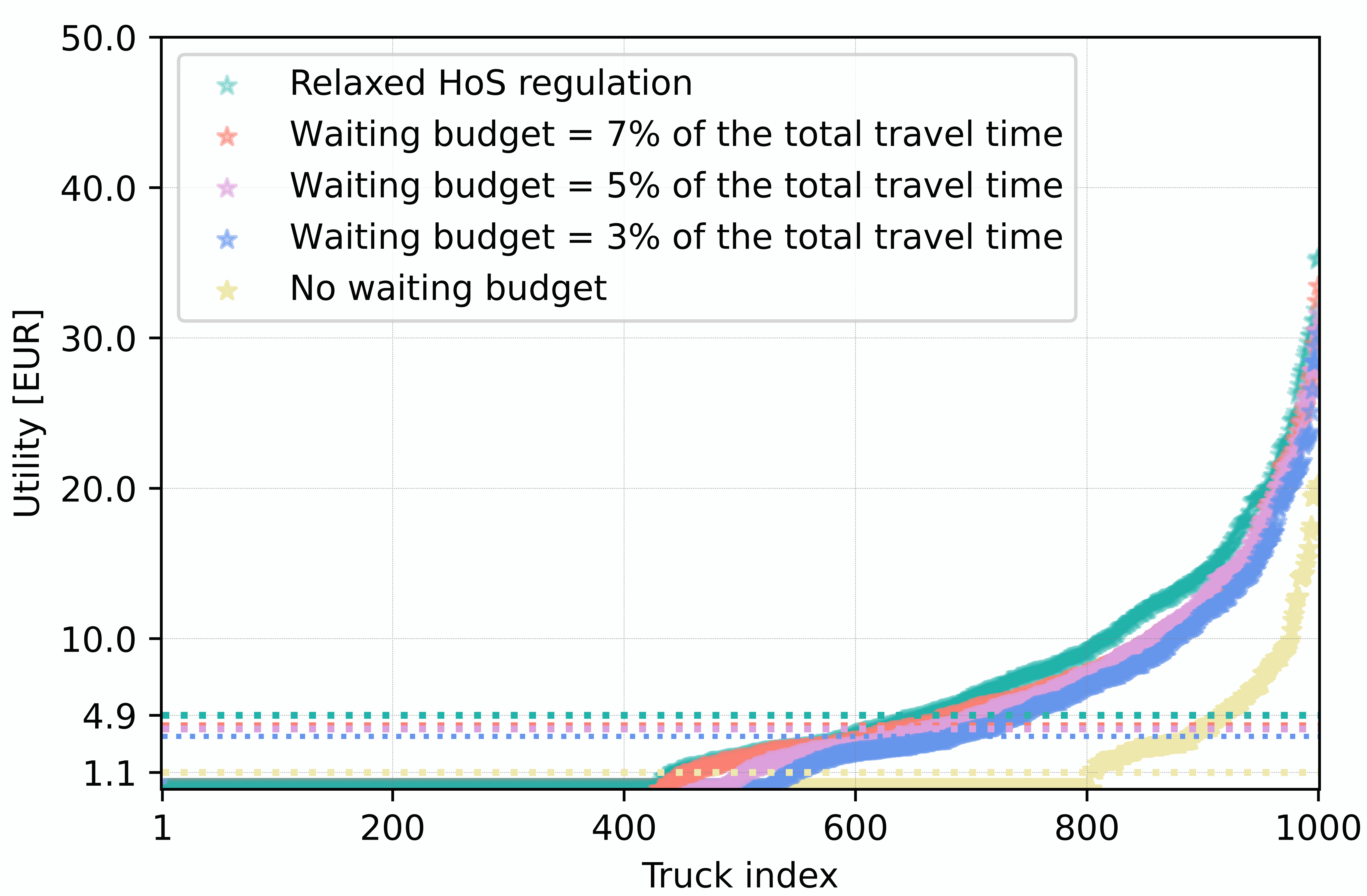}
     \vspace{-2pt}
      \caption{The utility of each truck and the average utility (dotted lines) for different waiting budgets.}
      \label{Fig.8}
\end{figure}

The achieved utilities of trucks under different constraint and waiting budget settings are shown in Fig.~\ref{Fig.8}. This figure shows that an average utility of $4.9$\texteuro ~is achieved when a relaxed HoS regulation is applied. In addition, the average utilities of trucks are $1.1$\texteuro, $3.5$\texteuro, $4.0$\texteuro, and $4.2$\texteuro ~for waiting budgets of $0~\%$, $3~\%$, $5~\%$ and $7~\%$ of the total travel times, respectively, which indicates that the utility generally increases with an increased waiting budget, as expected.

\section{Conclusions and Future Work}
This paper studied the problem of coordinating platoon formation at hubs while considering trucks' realistic HoS regulations. We developed a model of a transportation system, where trucks have fixed routes in the road network and optimize their platooning profits by deciding on their rest and wait times at hubs to respect HoS regulations. To solve the platoon coordination problem, an approximate DP approach was presented, where the decision-making of trucks is decoupled so that each truck can make decisions based on the predicted schedules of others. A large-scale simulation was conducted over the Swedish road network that considers the EU's HoS regulations. Our simulation results show that considerable platooning profits can be achieved under today's regulations and the waiting budget plays an import role for achieving a high platooning profit.   
 
The HoS regulations considered in this paper are slightly restrictive. For example, the regulations in the EU allow drivers to split the mandatory rest time ($45$ min) within one continuous driving period into $30$ min and $15$ min. In future work, we would like to extend the work in this paper to capture less restrictive rest time constraints.   

\section{Acknowledgment}
The authors would like to thank Albin Engholm for providing the simulation data from SAMGODS model, and the anonymous reviewers for their insightful comments. The author Ting Bai would also like to thank the support of the Career Development Scholarship from Shanghai Jiao Tong University.
\bibliographystyle{IEEEtran}
\bibliography{CDC_2022}
\end{document}